\documentclass[aps,prl,twocolumn]{revtex4}
\input epsf
\begin{document}
\title{Local electromigration model for crystal surfaces 
}
\author{O.\ Pierre-Louis}
\affiliation{CNRS/Laboratoire de Spectrom\'etrie Physique, UJF-Grenoble 1,
BP87, F38402 Saint Martin d'H\`eres, France}
\date{\today}

\begin{abstract}
We analyze the dynamics of crystal surfaces
in the presence of electromigration. 
From a phase field model with a migration force
which depends on the local geometry,
we derive a step model with additional contributions 
in the kinetic boundary conditions. 
These contributions trigger
various surface instabilities, such as step meandering,
bunching and pairing on vicinal surfaces. Experiments
are discussed.
\end{abstract}

\maketitle

In the presence of an electric current, 
mobile atoms experience a
diffusion bias, which is called electromigration.
Surface electromigration is known as major source of rupture
of microsctructures, and is also a interesting tool for
spontaneous nano-structure formation.
In the past 15 years, 
kinetic instabilities at the micro, or nanometer
scales, such as step bunching \cite{Latyshev1989},
meandering\cite{Yagi}, and pairing\cite{PierreLouis2004} on stepped
surfaces under electromigration have been
the focus of a large literature.
In the usual model for step motion under
electromigration\cite{Stoyanov1990}, 
mobile atoms between steps experience a
constant drift force, while
kinetic boundary conditions at steps
are  not affected by migration.
It was first suggested in Ref.\cite{PierreLouis200}
that migration could lead to additional terms
in step kinetic boundary conditions, and that these
terms could modify the electromigration-induced
drift of monolayer islands.
Nevertheless, a microscopic explanation
was lacking. 

From the work of Rous and Bly \cite{rous}
it is known that, in the vicinity of steps, the backscattering
of carriers strongly alters the migration force.
In the present letter, we show that
the boundary conditions used previously
in the literature should be modified 
to account for the variations
of the migration force in the step region.
To do so, we  derive a step model from
a phase field model with a non-constant migration force.
The analysis of this step model reveals that the new
contributions may produce all the instabilities 
known so far on vicinal surfaces.
We finally discuss the relevance of our
results for experimental observations
on various surfaces.

We start with a microscopic description
in terms of a phase field
model \cite{PierreLouis2003b}.  We use the simplest model,
where the surface is described by 
two fields $\phi$ and $c$, which
respectively represent the normalized 
height of the surface, and the local concentration
of mobile atoms. More sophisticated
models, which may account for arbitrary kinetics
at the steps are not considered here
for the sake of simplicity \cite{PierreLouis2003b}.
The dynamics of the phase field 
$\phi$ is such that $\phi$ relaxes
to a step and terrace structure. Terraces
are wide regions where $\phi$ is constant,
and steps have a typical width $W$, as shown in
Fig.1.
The motion of the steps is driven by the
departure from equilibrium in the step region,
measured by the concentration variations:
$(c-c_{eq})$, where $c_{eq}$
is the equilibrium concentration.
The dynamics of $\phi$ is a relaxation
with time-scale $\tau_\phi$:
\begin{eqnarray}
\tau_\phi\partial_t\phi=-\partial_\phi f+W^2\nabla^2\phi
+\lambda(c-c_{eq}) \partial_\phi g
\label{e:2f_pf}
\end{eqnarray}
The free energy density $f$ is a periodic
function of $\phi$ with minima for the values
of $\phi$ corresponding to the terraces.
The coupling function $g$ is also a 
function of $\phi$, with $\partial_\phi g>0$,
and  $g_--g_+=1$, where $\pm$ indicate
the lower or upper side of the step respectively.
The constant $\lambda$
controls the strength of the coupling
between mobile atoms and steps.

A global external force is present,
related to the macroscopic current
in the bulk of the crystal. This current induces
a local surface migration force ${\bf M}$ on mobile atoms.
The force ${\bf M}$ depends on the surface local geometry
via the phase field  $\phi$. The resulting 
mass flux, oriented in the direction ${\bf n}_{e}={\bf M}/|{\bf M}|$, reads:
\begin{eqnarray}
j_{e}=Dc {{\bf M} \over k_BT}={D c \over \xi}{\bf n}_{e}
\end{eqnarray}
where $D$ is the local $\phi$-dependent
diffusion constant, and $\xi=k_BT/|{\bf M}|$ is 
a $\phi$-dependent lengthscale which characterizes the 
amplitude of the force. Local mass conservation then reads:
\begin{eqnarray}
\partial_tc=\nabla\left[D
\left(\nabla c- {c \over \xi}{\bf n}_{e}\right)
\right]
+ F-{c \over \tau}-\partial_th
\label{e:2f_pf_diff}
\end{eqnarray}
where $h$ is the solid concentration, i.e. the
number of solid atoms per unit area above a
plane of arbitrary height parallel
to the terraces. The variations of $h$ 
are analogous to that of $\phi$, and
the jump of $h$ across a step is
constant $h_--h_+=1/\Omega$, where $\Omega$
is the atomic area.
Writing Eq.(\ref{e:2f_pf_diff}),
we have discarded the tensorial coupling between
the geometrical anisotropy of the step and the 
migration direction
(i.e. the orientation of ${\bf n}_{e}$), which could 
change the local migration direction in the
vicinity of a step.
We shall rather focus on another effect 
which is, in our opinion, the most important one:
the variation  of the
migration force amplitude in the step region.
The phase field model presented
here also neglects the dependence
of the migration force on the adatom density.
Indeed, this effect should occur at
high densities only \cite{Ishida2000},
while the adatoms densities which are 
relevant for the experiments mentioned below
are small.

\begin{figure}[h]
   \centerline{\epsfysize=8cm\epsfbox{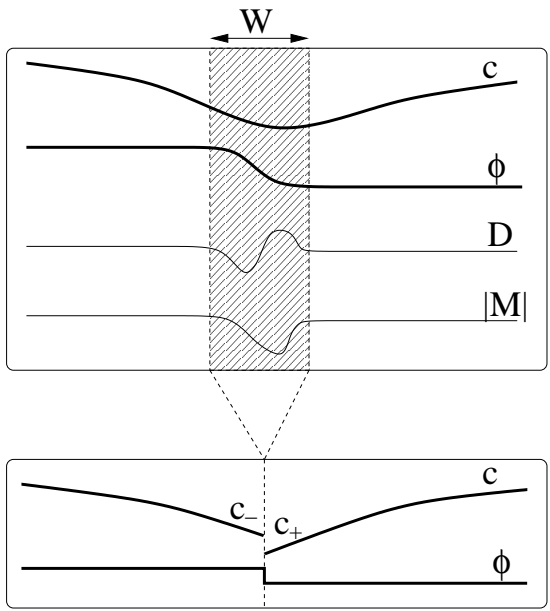}}
   \caption{Phase field model with a non-constant
migration force. The step region of width $W$
is hatched. For the sake of simplicity, we have shown a 1D cross-section,
but the model in the text is 2D. Notations:
$c$ is the mobile atom concentration,
$\phi$ the phase field, $D$ is the local diffusion constant, and 
$|{\bf M}|$ is the local amplitude of the migration force.
Using suitable asymptotics, one obtains a step model
as shown in the lower part of the figure, where the concentration
may be discontinuous at the step ($c_+\neq c_-$), and $\phi$ is now a Heaviside
function which indicates the step position. 
}
\label{fig1}
\end{figure}

We now derive a step model from the
phase field model (\ref{e:2f_pf},\ref{e:2f_pf_diff}).
When step kinetics is fast, one may perform
the thin interface asymptotics
reported in Ref.\cite{Karma,PierreLouis2003b}.
We shall assume that steps are slowly driven
out of equilibrium, so that the smallest cut-off 
length $\ell_c$ related to
the diffusion field on terraces is much larger than
the step width $W$. We therefore define a small parameter
$\epsilon\sim W/\ell_c$.
We also take the coupling constant
$\lambda\sim 1$, and the
departure from equilibrium $(c-c_{eq})\sim \epsilon$.
The latter condition indicates that step kinetics is fast enough
to keep the concentration in the step region close
to $c_{eq}$.  Furthermore, the relaxation dynamics of the phase 
field itself must be fast enough so that the
dynamics does not affect the
step and terrace structure, and we choose
$\tau_\phi\sim \epsilon^2$.

From these asymptotics, the dynamics of 
the concentration of atoms far from the steps
reduce to Eq.(\ref{e:2f_pf_diff}), with $D\rightarrow D_0$,
and  without the term $\partial_th$.

On the macroscopic level, and in the limit where
the adatom coverage is small, mass conservation at
the step reads:
\begin{eqnarray}
{V_n \over D_0\Omega}= {\bf n}.\nabla c_+
-{\bf n}.\nabla c_-
+  {c_--c_+\over \xi_s}
\label{e:mass_cons}
\end{eqnarray}
with
\begin{eqnarray}
{1 \over \xi_s}={{\bf n}. {\bf n}_{e} \over \xi_0}
\end{eqnarray}
where $\xi_0$ and $D_0$ are respectively 
the constant  value of $\xi$ and $D$
on terraces far from steps.
Performing the thin interface
asymptotics, we obtain Eq.(\ref{e:mass_cons}) to leading order.
But since $c\sim c_{eq}$ at
the steps to leading order, the last term
$(c_--c_+)/\xi_s$
only appears to higher orders.

The additional equations obtained from the thin interface
asymptotics are the kinetic boundary conditions
for the concentration at the steps\cite{tobepub}:
\begin{eqnarray}
D_0{\bf n}.\nabla c_\pm - {D_0 c_\pm \over \xi_s}
&=& {\nu}_\pm\left(c_\pm-\tilde c_{eq}-\beta {V_n \over \Omega}
\pm Q_\pm {c_{eq} \over\xi_s}\right)
\label{e:kin_bc}
\end{eqnarray}
where  $\tilde c_{eq}=c_{eq}(1+\Gamma\kappa)$.
These expressions involve some coefficients,
the definition of which follows:
\footnote{
We here ignore the terms related the interface
definition mentioned in Ref.\cite{PierreLouis2003b},
which are not relevant for our purposes.}
\begin{eqnarray}
\Gamma c_{eq} &=& 2^{1/2}{W \over \lambda}
\int d\phi \, (f-f_{min})^{1/2}
\nonumber \\
\beta &=& 2^{1/2}{\tau_\phi\over W} \int d\phi\, (f-f_{min})^{1/2}
\nonumber \\ && -
\int dz (g-g_-)(g_+-g)D^{-1}
\nonumber \\
\nu^{-1}_\pm &=&\pm \int dz (g_\mp- g)
\left( D^{-1}-D_0^{-1} \right)
\nonumber \\
Q_\pm &=& \pm \int dz (g_\mp- g)(1-\xi_0/\xi)
\label{e:Q_pf}
\end{eqnarray}
where $f_{min}$ is the value of $f$ at its minima.
Integrals are taken through one step in the
normal direction $z$.
The constants 
$g_\pm$ are the values of $g$ on both sides
of the steps.

If we ignore the terms proportional to $Q_\pm$,
Eqs.(\ref{e:kin_bc}) are equivalent
to the general linear kinetic boundary condition
of steps when step transparency
and attachment-detachment on both
sides of the steps are taken into account\cite{PierreLouis2003b}.
Using linearized  non-equilibrium
thermodynamics, Eqs.(\ref{e:kin_bc}) could have 
been stated directly, with the new terms 
proportional to $Q_\pm$
\footnote{ 
Indeed, the mass flux (l.h.s. of Eqs.(\ref{e:kin_bc}))
must be proportional to a linear combination
of the  3 thermodynamic
forces, which are proportional
to the departure from 
equilibrium on both sides of the step
$(c_+-c_{eq})$ and $(c_--c_{eq})$, and
the amplitude of the 
migration force far from the step $1/\xi_0$.
Then we follow Ref.\cite{PierreLouis2003b}
and perform a linear combination of the boundary conditions
to cancel the cross concentration terms, which introduces
the term $\beta V_n$.}.
Hence, Eqs.(\ref{e:kin_bc}) are
more general than 
the  specific limit of the specific phase field model
derived above. The relevance of our derivation from
a phase field model comes from the explicit link 
between the terms of Eqs.(\ref{e:kin_bc}) and
microscopic quantities.
Indeed, from Eqs.(\ref{e:Q_pf}), 
$Q_+$ or $Q_-$ account for
the variations of $|{\bf M}|$ on the lower
or upper side of the step respectively.
Moreover,  $Q_\pm>0$ or $Q_\pm<0$ respectively account for 
the a decrease or an increase of $|{\bf M}|$ in the step region.

Instead of analyzing the consequences
of the full boundary conditions (\ref{e:kin_bc}),
we shall now focus on a simplified limit.
We define:
\begin{eqnarray}
Q=Q_++Q_-=\int dx\, (1-\xi_0/\xi)\, .
\label{e:Q}
\end{eqnarray}
In order to grasp some of the main consequences 
of the new contributions, we shall focus on a
simplified limit where $Q_+=Q_-=Q/2$.
This may be obtained in the phase field model
with an antisymmetric $g$ and a symmetric variation
of $1/\xi$.  This assumption amounts to considering that 
the main effect of the force variation in the 
step region is an average increase, or
decrease of the  amplitude of the 
force, respectively leading
to a negative or positive $Q$. 
We also focus on a specific limit
for step kinetics. We define:
\begin{eqnarray}
d_0=D_0(\nu_+^{-1}+\nu_-^{-1})
=\int dx\,(D_0/D-1) .
\label{e:d_0}
\end{eqnarray}
Comparing the last term in the l.h.s.
with the last term in the r.h.s. of Eqs.(\ref{e:kin_bc}),
we infer that $Q_\pm$ will be relevant when
it is of the order magnitude as $D_0/\nu_\pm$.
We therefore take the limit where $\nu_\pm$ is large,
which implies $d_0$ small.
We also take $D$ to be symmetric
across the step so that $\nu_+=\nu_-$.

With these assumptions --symmetric $1/\xi$ and $D$
and fast kinetics-- the boundary conditions (\ref{e:kin_bc})  read:
\begin{eqnarray}
c_\pm=c_{eq}(1+\Gamma\kappa\mp Q^*/2\xi_s)+\beta V_n/\Omega
\label{e:kin_bc_transp}
\end{eqnarray}
where
\begin{eqnarray}
Q^*=Q+d_0 =\int dx \, \left({D_0 \over D}-{\xi_0 \over\xi}\right)
\label{e:Q*}
\end{eqnarray}
The expression (\ref{e:Q*})
indicates in a remarkably compact formulation,
that a single quantity $Q^*$ accounts at the same time for the variations
of the migration force, and for the variations 
of the diffusion constant in the step region.
Eqs.(\ref{e:kin_bc_transp}) account for transparent steps when
$\beta$ is large, and non-transparent steps
with fast attachment-detachment kinetics when
$\beta$ is small.

We shall now point out how  all known 
elementary instabilities of vicinal surfaces
are obtained from the model with kinetic boundary conditions
(\ref{e:kin_bc_transp}).
Let us  analyze the stability
of an array of initially straight and
parallel steps separated by the same distance
$\ell$.  We assume that the direction of migration
is orthogonal to the average step direction.
We also consider the conserved regime where
$F=0$ and $1/\tau\rightarrow 0$.
The $m$th step is perturbed 
by a small deviation 
\begin{eqnarray}
\zeta_m(x,t)=\zeta_{\omega\phi}\exp[i\omega t+im\phi+iqx]
\label{e:pertu}
\end{eqnarray}
where $\phi$, and $q$ are respectively 
the phase shift from step to step and 
the wavevector along the step. 
An instability is
indicated by a positive $\Re e[i\omega]$.
Substituting Eq.(\ref{e:pertu}) in the 
model equations, one finds the general
expression of $i\omega$ as a function of
$\phi$ and $q$.

In the case
of in phase meandering ($\phi=0$),
one finds in the long wavelength
and weak electromigration limit
($(q\ell)^2\sim \ell/\xi_0 \ll 1$):
\begin{eqnarray}
i\omega \approx \Omega c_{eq} D \left[ { Q^*\over \xi_0} q^2
-\Gamma\ell q^4\right]
\end{eqnarray}
Therefore, the train of steps is unstable
when $Q^*/\xi>0$, and the most unstable wavelength is:
$\lambda_m=2\pi (2\xi_0\Gamma\ell/Q^*)^{1/2}$.

In order to analyze step bunching, we shall
introduce a repulsion between steps. If this interaction
is of elastic or entropic origin, its free energy 
per unit step length is ${\cal A}/\ell^2$ \cite{noz}. 
The local equilibrium concentration
is then 
\begin{eqnarray}
\tilde c_{eq}=c_{eq}\left(1+\Gamma\kappa
+2A\left(\ell_+^{-3}-\ell_-^{-3}\right)\right)
\end{eqnarray}
where we have only considered 
the interaction between neighboring steps,
and $A=\Omega {\cal A}/k_BT$.
For the sake of simplicity, we analyze the
stability of the mode $q=0$, which accounts
for the bunching of straight steps.
In the limit of weak electromigration,
the linear stability  analysis leads to:
\begin{eqnarray}
{i\omega\over \Omega Dc_{eq}} =
{ (2Q^* /\xi_0 \ell^2) [1-\cos(\phi)]
-(12A /\ell^5) [1-\cos(\phi)]^2
\over
1+2[1-\cos(\phi)]\Omega D\beta/\ell}
\end{eqnarray}
Once again, an instability 
appears if $Q^*/\xi>0$. The stability
analysis thus indicates that 
a vicinal surface is simultaneously destabilized 
with respect to bunching and meandering for a downhill
flux when migration or diffusion is weaker in the step
region, and for an uphill migration
when migration or diffusion is enhanced 
in the vicinity of the steps.

Let us now discuss some of the experimental
results of the literature. On Si(111),
atoms drift along the
current direction \cite{minoda}.   
Three temperature regimes, denoted I,II, and II, with increasing
temperature, were found above
the $(7\times 7)\rightarrow (1\times 1)$ reconstruction
transition temperature \cite{Latyshev1989,metois}.
In regimes I and III,
bunching is observed for a downhill
current during both growth and sublimation\cite{metois}.
In regime II, bunching is observed during growth
for a downhill current,
and during sublimation for an uphill
current \cite{metois}. 
From the step model without the $Q$ contribution,
it was concluded that regimes I and III
correspond to opaque steps ($\beta$, $\nu_+$, and $\nu_-$ small),
and regime II corresponds to fast step kinetics
($\nu_+$, and $\nu_-$ large) \cite{metois,PierreLouis2003a}.

In the case of opaque steps, the $Q$ terms
are probably only a small contribution, unless a dramatic
increase of migration is observed in a large
region around the step. There is no evidence
of such a strong effect. 

In the case of fast kinetics,
the $Q$ terms become important.
From the study of
step pairing \cite{PierreLouis2004}, it was concluded that 
$\beta$ is large, which means that
steps are transparent.
In Ref.\cite{PierreLouis2004}, the occurrence
of pairing was related to a small negative
kinetic length $d_0\approx -0.13{\rm \AA}$.
In the present analysis, this result generalizes to 
$Q^*=Q+d_0 \approx -0.13{\rm \AA}$.
We therefore obtain
an alternative explanation for the pairing
in the limit $|Q|\gg|d_0|$ of steps on Si(111), 
based on $Q<0$, i.e. on a stronger migration in the step regions.
Quantitatively, we find $Q\approx -0.13{\rm \AA}$.
This may for example account for an increase
of $1\%$ of the migration force 
in a step region  of width $13 {\rm \AA}$.

Step meandering
is also observed during sublimation with
downhill migration in regime II \cite{Yagi}.  If $Q<0$, as suggested above, then
$Q$ rather favors the stabilization
of the meander for a downhill flux.
Meandering therefore requires an extended
study, which would also account for the sublimation rate,
and  which is beyond the scope
of the present letter. A recent discussion
can be found in Ref.\cite{Tong2004}.

Another system of interest is Si(100), which
undergoes a dimer-row reconstruction,
rotated of $\pi/2$ from one layer to the next one.
We here assume
that dimer rows are perpendicular $\bot$ or 
parallel $\parallel$ to  steps,
and that the current is perpendicular to the average
step orientation. It was shown in Ref.\cite{Stoyanov1990}
that the difference of diffusion constant ($D_\bot$, 
$D_\parallel$) perpendicularly
to the step leads to pairing,
and subsequent bunching of pairs \cite{sato2004}
for both directions of the electric current. 
The crucial quantity on terraces
is the migration induced mass flux $j \sim c_{eq} D/\xi$, 
which is proportional to the product of the diffusion 
constant with the migration force. Therefore,
the phenomena that are attributed to
a variation of $D$ can also be the result
of a variation of $\xi$. Thus, pairing and bunching
of pairs can also be the result variations of $\xi$.
More precisely, stronger migration
along dimer rows reproduces pairing and bunching
of pairs for both directions of the current,
as observed in experiments \cite{Litvin1991}.
As a conclusion for Si(100),
the variations of the force
cannot be neglected a priori.
Furthermore, following Ref.\cite{Tong2005}, one could consider
pairs  as effective steps,
with an internal region where diffusion
and the migration force may vary. We could then
follow the above mentioned analysis to
determine the effective value of $Q$.
This would help to analyze the pair bunching
dynamics.

We also expect the effect presented above
to be relevant for the case of metals.
Indeed, from microscopic models on
metallic surfaces \cite{rous}, strong
variations of the migration force in the 
vicinity of the steps were found.
Rous and Bly \cite{rous},
indicate a decrease up to
$\sim 50\%$ in a region $\sim 20 {\rm \AA}$ for Na.
Using Eq.(\ref{e:Q}) with an approximate
integration of the results of Ref.\cite{rous}, we find
$Q\approx 10 {\rm \AA}$.
This contribution is two order of magnitude larger
than that found above for Si(111),
and should therefore have drastic consequences
on the dynamics of metal surfaces
under electromigration.
Experiments on Au surfaces \cite{Shimoni1997} indicate
that in presence of an electric current,
the Ehrlich-Schwoebel (ES) effect \cite{Schwoebel1969}
(a decrease of interlayer mass transport related
to a low value of $\nu_-$) disappears, and a significant
increase of interlayer mass transport is observed.
The disappearance of the ES
effect is probably caused by the increase of the
temperature due to the joule heating effect.
Indeed, it was shown in Ref.\cite{Ferrando1996}, that the
ES effect disappears at high temperature on Au.
Therefore, kinetics is fast, 
and the tendency to bunching indicated in Ref.\cite{Shimoni1997}
could be caused by the non-constant migration
scenario presented above.

Step bunching was also observed on some
other metal surfaces \cite{Johnson1938}.
Moreover, recent experiments \cite{Xie2000} have shown that GaN(0001)
vicinal surfaces also exhibit step bunching
for a downhill electric current during epitaxial
growth. The non-constant migration force scenario
should also be considered in these cases.
But the lack of quantitative understanding of the microscopic
processes prevents a precise
analysis.

In conclusion, we have shown that 
variations of the migration force
in the vicinity of the steps
may lead to step bunching, meandering, or pairing
during electromigration. This mechanism
defines a novel scenario for surface instabilities,
which may compete or combine with
the other destabilizing mechanisms 
analyzed in the literature.
Finally, we have 
discussed the relevance of our results
for some semiconductor and metal surfaces.

\end{document}